\documentclass[11pt]{article}
\usepackage{moriond,epsfig}
\usepackage{graphicx}
\usepackage[figuresright]{rotating}
\usepackage{pstricks}
\usepackage{amsmath}

\newcommand{\lbl}[1]{\label{eq:#1}}

\newcommand{\be}{\begin{equation}}
\newcommand{\ee}{\end{equation}}
\newcommand{\bea}{\begin{eqnarray}}
\newcommand{\eea}{\end{eqnarray}}

\newcommand{\ra}{\rightarrow}

\newcommand{\lesssim}{ {\
\lower-1.2pt\vbox{\hbox{\rlap{$<$}\lower5pt\vbox{\hbox{$\sim$}}}}\ } 
}
\newcommand{\gtrsim}{ {\
\lower-1.2pt\vbox{\hbox{\rlap{$>$}\lower5pt\vbox{\hbox{$\sim$}}}}\ } 
}

\newcommand{\cO}{{\cal O}}

\newcommand{\Imm}{\mbox{\rm Im}}

\newcommand{\MeV}{\mbox{\rm MeV}}

\newcommand{\annd}{\mbox{\rm and}}

\newcommand{\bfw}{\mbox{\bf w}}

\begin{document}

\title{Long-distance contributions in $ K \rightarrow \pi l^+ l^-$ Decays}

\author{Samuel FRIOT$^{a, b}$ and David GREYNAT$^a$}

\vspace*{0.5cm}

\address{~\\$^a$Centre  de Physique Th\'eorique \\Case 907, 13288 Marseille Cedex 9, France \\CNRS UMR 6207}

\address{$^b$Laboratoire de Math\'ematiques Universit\'e Paris-Sud \\B\^at. 425 , 91405 Orsay Cedex, France \\CNRS UMR 8628}

\maketitle\abstracts{
This talk gives a short summary of some work~\cite{FGEdeR04} realized in collaboration with E. de Rafael and concerning $K\ra\pi \ell^+ \ell^-$ decays, in a combined framework of Chiral Perturbation Theory and  Large--$N_c$ QCD under the dominance of a minimal narrow resonance structure.} 

\section{Introduction}\lbl{int}

Nowadays, rare K decays know a particularly increasing interest since the branching ratios of some still unobserved of them, that are golden modes for CP--violation study, are predicted to be rather closed to actual experimental limits. Two of the most known and most promising processes belonging to this class are $K_L\ra\pi^0 \nu \bar\nu$ and $K_L\ra\pi^0 e^+ e^-$.

In ~\cite{FGEdeR04}, one of the aims was the evaluation of the $K_{L}\ra \pi^{0}e^{+}e^{-}$ long-distance contributions. Until 2003 and the first observation and  precise measurement by the NA48 Collaboration at CERN~\cite{NA48} of the modes $K_{S}\ra \pi^{0}e^{+}e^{-}$ and $K_{L}\ra \pi^0 \gamma \gamma$, no really reliable theoretical estimate of the $K_{L}\ra \pi^{0}e^{+}e^{-}$ branching ratio was available, since this quantity was thought, at least from some viewpoint, to depend crucially on these two decays.
Then, after these new measurements, a theoretical re-analysis of this golden mode was obtained in ~\cite{BDI03}. Our work ~\cite{FGEdeR04} propose an alternative strategy for part of the theoretical analysis elaborated in this last paper (and in the correlated paper ~\cite{DEIP98}). 
We shall not dive here into theoretical details that can be found in our original publication (see also ~\cite{G04}) but instead insist on some of the underlying ideas.

Apart from its CP--conserving part which is known to be negligible, the branching ratio of the $K_{L}\ra \pi^{0}e^{+}e^{-}$ process fragments into three parts: a direct CP--violating (CPV) part, an indirect one (also called mixing) and the important interference between them,
\begin{align}
\left.{\rm Br}\left(K_L \rightarrow \pi^0 e^+ e^-\right)\right\vert_{\rm\tiny CPV}&= \left[C_{dir.} \left(\frac{{\rm Im}\lambda_t}{10^{-4}}\right)^{2}+C_{ind.} \pm C_{int.}\, \frac{{\rm Im}\lambda_t}{10^{-4}}\right]\times 10^{-12}\label{KSCPV}\,	,
\end{align}
where $C_{dir.}=2.4\pm0.2$ is known precisely due to its (perturbative) short-distance origin, $C_{ind.}=10^{12}|\epsilon|^2\frac{\tau(K_L)}{\tau(K_S)}{\rm Br}(K_S\rightarrow\pi^0e^+e^-)$ and $C_{int.}\sim\sqrt{C_{dir.}C_{ind.}}$\; (see ~\cite{BDI03}).

With NA48 results concerning ${\rm Br}(K_S\rightarrow\pi^0e^+e^-)$, we can directly conclude that the indirect CPV contribution is quite large.
This apparently overwhelming CP violating behaviour of $K_L \rightarrow \pi^0 e^+ e^-$ explains the importance of this decay mode and thus one major question concerns the type of interference (constructive or destructive, depending on the algebraic value of $C_{int.}$). Arguments in favour of a constructive interference have been suggested in ~\cite{BDI03} and are good news for future experimental investigations; as we shall see in the following, we found the same conclusion within the approach that we suggested. 
 
\section{Indirect CPV and interference contributions: $\mathcal{O}(p^4)$ Chiral Perturbation theory and beyond}
 
In our work, two of the main points of study are the indirect CPV and interference contributions of ${\rm Br}(K_L\rightarrow\pi^0e^+e^-)$. We wish to emphasize that contrary to ~\cite{BDI03} who rely on the $K_S\rightarrow\pi^0e^+e^-$ channel for prediction of the indirect CPV term, we showed in ~\cite{FGEdeR04} that both contributions can be related to the much more experimentally known charged mode $K^+\rightarrow\pi^+e^+e^-$ for which our approach gives also a description consistent with all corresponding available data. It results from this viewpoint a semi-theoretical prediction for the indirect CPV contribution competitive to the purely experimental deduction of ~\cite{BDI03}. Moreover, our approach, by its common framework for the neutral and charged modes, has less free parameters than what the authors of ~\cite{BDI03} had to introduce in order to obtain at the same time their conclusion for the interference term and a theoretical description of the charged mode compatible with experiments (see also ~\cite{DEIP98} for part of the original work concerning the latter).

The low-energy description of $K\ra \pi\ell^+\ell^-$ decays is based on Chiral Perturbation Theory ($\chi$PT) and, for the specific case of $K_{L}\ra \pi^{0}e^{+}e^{-}$, the CPV--branching ratio (\ref{KSCPV}) can be rewritten, at $\mathcal{O}(p^4)$ of $\chi$PT, as \footnote{This equation is obtained by keeping only $\mathcal{O}(p^4)$ terms and neglecting constant and linear terms in $a_s$ in eq. (30) of ~\cite{BDI03}.}
\begin{align}
{\rm Br}\left(K_L \rightarrow \pi^0 e^+ e^-\right)\vert_{\rm\tiny CPV}&= \left[(2.4\pm 0.2) \left(\frac{{\rm Im}\lambda_t}{10^{-4}}\right)^{2}+(3.5\pm 0.1)\,\left(\frac{1}{3}-{\bfw}_s\right)^2 \right. \nonumber\\
&\hspace{4cm}\left.+ (2.9\pm 0.2)\, \left(\frac{1}{3}-{\bfw}_s\right)\frac{{\rm Im}\lambda_t}{10^{-4}}\right]\times 10^{-12}\label{KSCPV2}\,	.
\end{align}
As can be seen in this formula, the type of interference is directly correlated to one unknown constant ${\bf w}_s$ that we have to evaluate (the other constant $\Imm\lambda_t$ is known to be $(1.36\pm 0.12)\times 10^{-4}$, see ~\cite{Baetal03}).
With (\ref{KSCPV2}) in mind, we can now basically explain the strategy that has been proposed in our work in order to obtain a prediction for this branching ratio but in this respect, we have to tell more about ${\bf w}_s$.

In fact, ${\bf w}_s$ is mainly the combination of low-energy chiral coupling constants that appear in the renormalized $\mathcal{O}(p^4)$ $\chi$PT decay rate's expression of the $K_{S}\ra \pi^{0}e^{+}e^{-}$ neutral mode ~\cite{EPR87,FGEdeR04},
\begin{equation}
\label{ws}
{\bf w}_s=-\frac{1}{3}(4\pi)^2\;\tilde{\bf w} -\frac{1}{3}\log\frac{M_K^2}{\nu^2}\;,
\end{equation}
where $\tilde {\bf w}={\bf w}_1-{\bf w}_2$.
It is possible to have a completely theoretical estimate of these coupling constants by finding their underlying Green's functions that can then be evaluated in the Large-$N_c$ expansion of QCD (see, for an example of such procedure, ref. ~\cite{HPR03}). We first followed another, more phenomenological, approach. 

One should notice that, up to the logarithm, the same coupling constants combination also appears in the $\mathcal{O}(p^4)$ $\chi$PT expression of the $K^+\ra\pi^+ e^+ e^-$ decay rate, in the corresponding ${\bf w}_+$ constant ~\cite{EPR87,FGEdeR04}
\begin{equation}
{\bf w}_+=-\frac{1}{3}(4\pi)^2\;\tilde{\bf w} -(4\pi)^2\;[{\bf w}_2-4{\bf L}_9]-\frac{1}{6}\log\frac{M_K^2m_\pi^2}{\nu^4}\;.
\end{equation}
This is one of the two important points that make that the prediction for the indirect CPV and interference terms of ${\rm Br}(K_L\rightarrow\pi^0e^+e^-)$ can be obtained entirely from the charged mode (the second point is that our "beyond $\mathcal{O}(p^4)$" $\chi$PT approach does not introduce more free parameters than what "pure $\mathcal{O}(p^4)$" $\chi$PT does, so that our unique unknowns in the "beyond $\mathcal{O}(p^4)$" $\chi$PT approach are still $\tilde{\bf w}$ and ${\bf w}_2-4{\bf L}_9$). Now, this is precisely the charged mode which is at the origin of the  completely different theoretical analyses done in ~\cite{BDI03} and in our work.

In fact, two experimental informations are available for the charged decay: its branching ratio and the mass spectrum ~\cite{zeller}. Now, $\chi$PT at $\mathcal{O}(p^4)$ fails to correctly reproduce the slope of the mass spectrum. Two different strategies can then be followed to solve this problem. The first one ~\cite{DEIP98} (see also ~\cite{I01}) is to consider that $\mathcal{O}(p^6)$ corrections are important and that it is necessary to go to this order to have a good description of the form factor that enters into the theoretical expression of the $K^+\ra\pi^+ e^+ e^-$ decay rate. 
This is what the authors of ~\cite{DEIP98} have done in an approximate way for the charged channel (and as a by product for the neutral case) since the $\mathcal{O}(p^6)$ $\chi$PT analysis is too complicated to be done without approximation. With their approach, they were able to fit the $K^+\ra\pi^+ e^+ e^-$ mass spectrum and their analysis of the neutral mode has then been reinvestigated in ~\cite{BDI03} under the light of the NA48 new measurements, from which it has been possible to obtain a prediction for the indirect CPV and interference parts of ${\rm Br}(K_{L}\ra \pi^{0}e^{+}e^{-})$.
The second strategy to solve the $\mathcal{O}(p^4)$ problem is the one that has been put forward in our work. We proposed in ~\cite{FGEdeR04}, with the help of a Large-$N_c$ QCD inspired model, a second way of dealing with the $\mathcal{O}(p^4)$ problem by slighty modifying the $\mathcal{O}(p^4)$ form factor expression with an explicit replacement of the $\mathcal{O}(p^4)$ couplings by their underlying minimal resonances structure. 
This achieves a resummation to all orders of $\chi$PT which then goes beyond $\mathcal{O}(p^6)$. More precisely, we replaced the 
$\cO(p^4)$ form factor ~\cite{EPR87,FGEdeR04}
\begin{equation}
\label{fff1}
f_V(z)  = \frac{G_8}{G_F} \left\{ \frac{1}{3}-{\bf w}_+ - \frac{1}{60}z - \chi(z) \right\}\;,
\end{equation}
with $\chi(z)=\phi_{\pi}(z)-\phi_{\pi}(0)$ and  $\phi_{\pi}(z)=-\frac{4}{3}\frac{m_\pi^2}{M_K^2z}+\frac{5}{18}+\frac{1}{3}\left(\frac{4m_\pi^2}{M_K^2z}-1\right)^{\frac{3}{2}}\arctan{\left(\frac{4m_\pi^2}{M_K^2z}-1\right)^{-\frac{1}{2}}}$, by the formula
\begin{align}
\label{fff2}
f_V(z) &= \frac{G_8}{G_F} \left\{  \frac{(4\pi)^2}{3} \left[{\tilde{\bf w}}\frac{M_{\rho}^2}{M_{\rho}^2-M_{K}^2 z} + 6   F_{\pi}^2 \beta \frac{M_\rho^2 - M_{K^*}^2}{\left(M_\rho^2 - M_K^2 z\right)\left(M_{K^*}^2-M_K^2 z\right)}\right] \right. \nonumber\\
&\hspace{6cm}\left. + \frac{1}{6} \ln \left(\frac{M_K^2 m_\pi^2}{M_\rho^4}\right) +\frac{1}{3} - \frac{1}{60}z - \chi(z) \right\}\,,
\end{align}
where $\beta=-\frac{M_\rho^2}{2F_\pi^2}(1-\frac{M_\rho^2}{M_{K^*}^2})^{-1}({\bf w}_2-4{\bf L}_9)$. We kept the lowest order chiral loop contribution as the leading manifestation of the Goldstone dynamics.
As one can notice, no new free parameter appears in (\ref{fff2}) compared to (\ref{fff1}).

Now, with $\tilde {\bf w}$ and $\beta$ left as free parameters, we can make a non-linear regression to the mass spectrum data of ~\cite{zeller}. This is a typical errors-in-variables statistical problem but, in a first approximation, we neglected the invariant mass error of the data and the particles masses errors appearing in (\ref{fff2}), planning to take into account these effects in a future statistical study that will also include the not yet available but forthcoming NA48 mass spectrum data.
\begin{figure}[h]
\begin{center}
\includegraphics[width=0.7\textwidth]{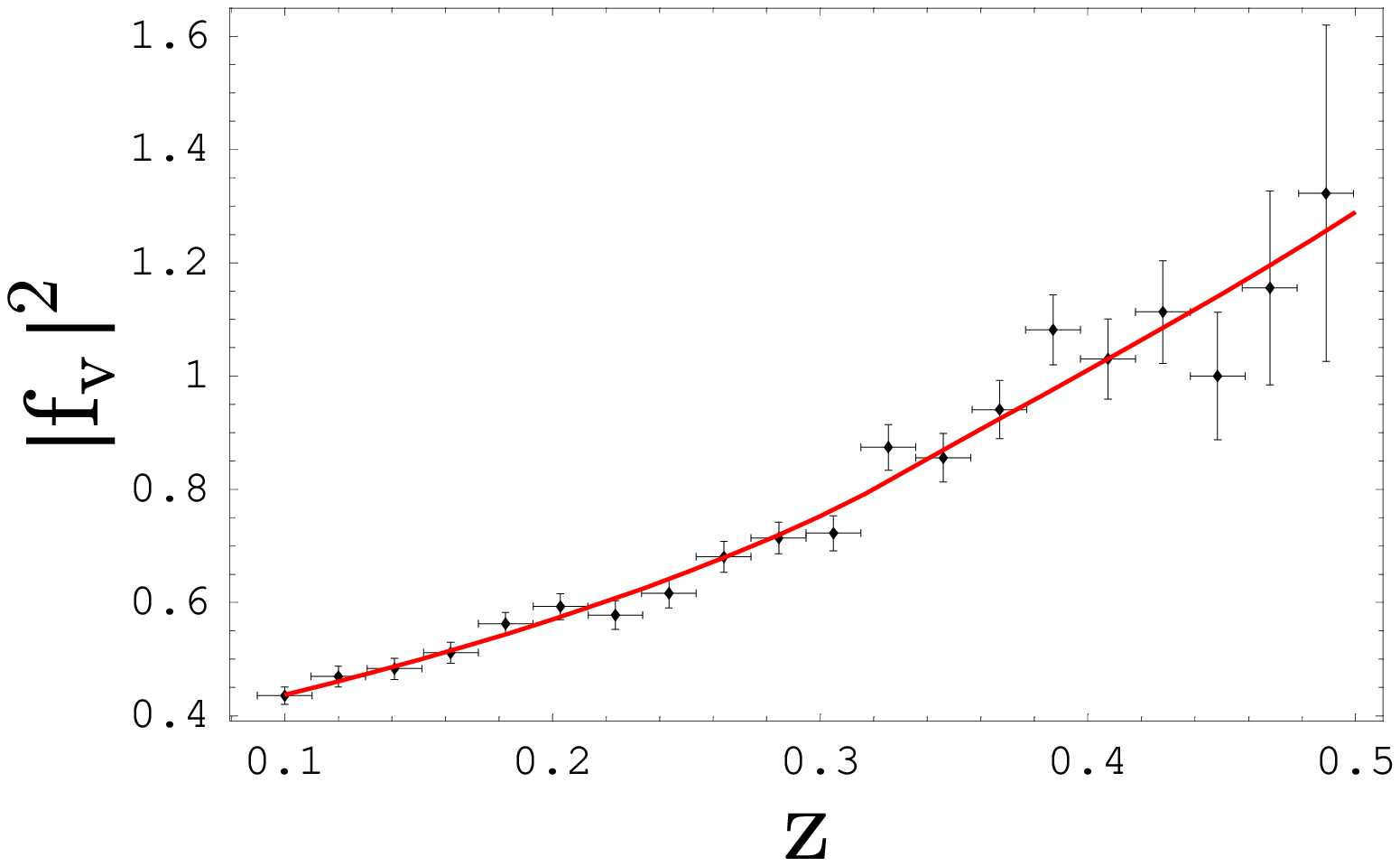}
\end{center}

{\bf Fig.~1} {\it Plot of the form factor $\left|f_V(z)\right|^2$ versus the invariant mass squared of the $e^+ e^-$ pair normalized to $M_{K}^2$. The crosses are the experimental points ~\cite{zeller}; the continuous line is the fit of the form factor in eq.(\ref{fff2}).}
\end{figure}
The result is the continuous curve shown in the Figure 1, which corresponds to a $\chi_{\mbox{\rm\tiny min.}}^2=13.0$ for 18 degrees of freedom. The fitted values of the parameters (using ${\bf g}_8=3.3$ and $F_{\pi}=92.4~\MeV$) are 
\begin{equation}\label{wtildepred}
	\tilde {\bf w}=0.045\pm 0.003\qquad\annd\qquad {\bf \beta}=2.8\pm 0.1\,.
\end{equation}
We then compute the $K^+\ra\pi^+ e^+ e^-$ branching ratio, using (\ref{fff2}) and the fitted values for $\tilde {\bf w}$ and $\beta$, with the result
\begin{equation}
{\rm Br}(K^+\ra\pi^+ e^+ e^-)=(3.0\pm 1.1)\times 10^{-7}\,,
\end{equation}
in good agreement with experiment result
\begin{equation}
{\rm Br}(K^+\ra\pi^+ e^+ e^-)=(2.88\pm 0.13)\times 10^{-7}\,.
\end{equation}
A form factor similar to (\ref{fff2}) can be obtained for the case of ${\rm Br}(K_{S}\ra \pi^{0}e^{+}e^{-})$ and making use of (\ref{wtildepred}), we find
\bea
\label{Kshort}
		{\rm Br}(K_S \rightarrow \pi^0 e^+ e^-) & = &  (7.7\pm 1.0)\times 10^{-9}\, 
\eea
and
\bea
{\rm Br}(K_S \rightarrow \pi^0 e^+ e^-)\vert_{>165{\scriptsize \MeV}} & = & (4.3\pm 0.6)\times 10^{-9} \,.
\eea
This is to be compared with the NA48 results ~\cite{NA48} 
\begin{equation}\lbl{K0rate}
	{\rm Br}(K_S \rightarrow \pi^0 e^+ e^-)= \left[5.8^{+2.8}_{-2.3} (\rm stat.) \pm 0.8 (\rm syst.) \right] \times 10^{-9}\,,
\end{equation}
and
\begin{equation}
{\rm Br}(K_S \rightarrow \pi^0 e^+ e^-)\vert_{>165{\scriptsize \MeV}}=		\left[3^{+1.5}_{-1.2} (\rm stat.) \pm 0.1 (\rm syst.) \right] \times 10^{-9}\,.
\end{equation}

The predicted branching ratios for the $K\ra\pi~\mu^{+}\mu^{-}$ modes are
\begin{equation}
	{\rm Br}(K^+ \rightarrow \pi^+ \mu^+ \mu^-)  =   (8.7\pm 2.8)\times 10^{-8} \quad\annd\quad	{\rm Br}(K_S \rightarrow \pi^0 \mu^+ \mu^-)  =   (1.7\pm 0.2)\times 10^{-9}\,,	
\end{equation}
to be compared with
\bea
			{\rm Br}(K^+ \rightarrow \pi^+ \mu^+ \mu^-)  & = &  (7.6\pm 2.1)\times 10^{-8}\,, \quad {\mbox{\rm ref.~\cite{K+dacrate}}} 
\eea
and
\bea
{\rm Br}(K_S \rightarrow \pi^0 \mu^+ \mu^-)  & = & \left[2.9^{+1.4}_{-1.2} (\rm stat.) \pm 0.2 (\rm syst.) \right]\times 10^{-9}\,, \quad {\mbox{\rm ref.~\cite{Moriond}}} .
\eea

Finally, the resulting negative value ${\bf w}_s=-2.1\pm0.2$ in (\ref{ws}) implies
 a constructive interference in the "pure $\mathcal{O}(p^4)$" $\chi$PT expression (\ref{KSCPV2}) with a predicted branching ratio
\begin{equation}\label{KLCPVP}
		{\rm Br}(K_L \rightarrow \pi^0 e^+ e^-)\vert_{\rm\tiny CPV}=(3.4\pm 0.4)\times 10^{-11}\,, 
\end{equation}
where we have used~\cite{Baetal03} $\Imm\lambda_t= (1.36\pm 0.12)\times 10^{-4}$.

When taking into account the effect of the modulating form factor by using (\ref{KSCPV}) and (\ref{Kshort}), one finds
\begin{equation}\label{KLCPVP2}
		{\rm Br}(K_L \rightarrow \pi^0 e^+ e^-)\vert_{\rm\tiny CPV}=(3.8\pm 0.4)\times 10^{-11}\,, 
\end{equation}
which indicates that higher order terms in the chiral expansion seem not too important.

\section{Conclusions}

Earlier analyses of $K\ra\pi~e^+ e^-$ decays within the framework of $\chi$PT have been extended beyond the predictions of $\cO(p^4)$, by replacing the local couplings which appear at that order by their underlying  narrow resonance structure in the spirit of the MHA to Large-$N_c$ QCD. The resulting modification of the $\cO(p^4)$ form factor is very simple and does not add new free parameters. Taking as input the invariant $e^+ e^-$ mass spectrum only, our strategy allows to reproduce, within errors, all known $K\ra\pi \ell^+ \ell^-$ branching ratios. The predicted interference between the {\it direct} and {\it indirect} CP--violation amplitudes in $K_L\ra\pi^0 e^+ e^-$ is constructive, with an expected branching ratio (see (\ref{KLCPVP2})) within reach of a dedicated experiment.

\section{Acknowledgments}

We are grateful to Eduardo de Rafael who is at the origin of this work and for his remarks concerning the manuscript. We are also grateful to J\'er\^ome Charles for his help in the beginning implementation of our future statistical analysis. One of the authors, D.G., expresses his gratitude to Antonio Pich for the invitation to the Moriond Conference.

\end{document}